\documentclass[preprint,3p,twocolumn,times]{elsarticle}

\usepackage{amsmath}
\usepackage{amssymb}
\usepackage{bm}
\usepackage{graphicx}
\usepackage{epsfig}
\usepackage{hyperref}
\usepackage[english]{babel}
\usepackage{color}

\biboptions{longnamesfirst,comma,square,numbers}

\journal{Solid State Communications}

\begin{document}

\begin{frontmatter}

\title{Antiferromagnetic Phase Diagram of the Cuprate Superconductors}

\author[UFSJ]{L. H. C. M. Nunes}
\ead{lizardonunes@ufsj.edu.br}
\author[UFSJ]{A. W. Teixeira}
\author[UFRJ]{E. C. Marino}
\ead{marino@if.ufrj.br}

\address[UFRJ]
{Instituto de F\'{\i}sica, Universidade Federal do Rio de Janeiro, Caixa Postal 68528, Rio de Janeiro, RJ, 21941-972, Brazil}

\address[UFSJ]
{Departamento de Ci\^encias Naturais, Universidade Federal de S\~ao Jo\~ao del Rei, 36301-000 S\~ao Jo\~ao del Rei, MG, Brazil}

\begin{abstract}
Taking the spin-fermion model as the starting point for describing the cuprate superconductors,
we obtain an effective nonlinear sigma-field hamiltonian,
which takes into account the effect of doping in the system.
We obtain
an expression
for the spin-wave velocity as a function of the chemical potential.
For appropriate values of the parameters
we determine the antiferromagnetic phase diagram
for the YBa$_2 $Cu$_3 $O$_{ 6 + x } $ compound
as a function of the dopant concentration
in good agreement
with the experimental data.
Furthermore,
our approach provides a unified description
for the phase diagrams of the hole-doped and the electron doped compounds,
which is consistent with the remarkable similarity
between the phase diagrams of these compounds,
since we have obtained the suppression of the antiferromagnetic phase
as the modulus of the chemical potential increases.
The aforementioned result then follows by
considering positive values of the chemical potential
related to the addition of holes to the system,
while negative values correspond to the addition of electrons.
\end{abstract}


\end{frontmatter}

\section{Introduction}
\label{introduction}

Cuprates are puzzling materials;
the undoped parent compounds are Mott insulators
presenting an antiferromagnetic (AF) arrangement at a finite N\'eel temperature.
As the system is doped, either with electron acceptors or donors,
which, in any case would simply mean an increase of the amount of charge carriers to the system,
it develops high-temperature superconductivity.
The superconducting (SC) critical temperature
presents a characteristic dome-shaped dependence on dopant concentration,
reaching a maximum value at an optimal doping
and vanishing as the system is doped even further,
becoming a normal metal.
So far, there is no consensus regarding the microscopic mechanism
which is responsible for the appearance of superconductivity in those systems.
However, it is widely accepted
that some sort of AF spin fluctuations
are the interaction responsible for the Cooper pairs formation.

Recently, starting from a spin-fermion model,
which has been employed  previously
to describe the cuprate superconductors
~\cite{Kotliar1990}
we have
derived an effective model for the charge carriers~\cite{AOP}
and the
superconductivity in our model
arises from a novel mechanism,
%
that yields a high critical temperature
which is comparable to the experimental values.
Moreover, by including doping effects
a dome-shaped dependence of the critical temperature
is found as charge carriers are added to the system,
in agreement with the experimental phase diagram~\cite{EPL}.
Presently
instead of focusing on the SC phase,
we investigate the magnetic order,
by calculating the doping dependence of the N\'eel temperature,
%
which is calculated
providing the AF phase diagram
that can be compared with experimental results.
As shall be seen below,
our results are in good agreement
with the data for the
YBa$_2 $Cu$_3 $O$_{ 6 + x } $
(YBCO) system~\cite{Rybicki2016, Rossat-Mignod1991}.
This calculation is an alternative to the one performed before \cite{Marino2001},
which was based on the fact that a skyrmion topological excitation is created
in association to a doped charge, being actually attached to it.

Our approach provides a unified description
for the phase diagrams of the hole-doped and the electron doped compounds,
which is consistent with the experimental results~\cite{Armitage2010}.
As the dopant concentration increases,
a dome-shaped SC phase appears adjacent to the AF order, both
for electron-doped and hole-doped cuprate compounds as well.
Our calculations present the suppression of the AF phase
as the modulus of the chemical potential increases.
Positive values of the chemical potential
are related to the addition of holes to the system,
while negative values correspond
to the addition of electrons.

\section{The spin-fermion model}
\label{SFModel}

So far, there is no consensus regarding the minimal model
which entails the vast phenomenology presented
by the high-$ T_c $ superconductors,
however,
since it is well established that
superconductivity emerges
from the CuO$_2 $ planes of the cuprates,
the paradigmatic three-band Hubbard model
proposed by Emery~\cite{Emery}
is a good candidate to describe the physics in these planes.
However, due to its several parameters
this model is a complicated starting model
for the study of the electronic properties in the  CuO$_2 $ planes.
Therefore, many theoretical studies have considered instead
a one-band Hubbard model
or a single band $t$-$J$ model~\cite{tJModel1}
which represents the lower ``Zhang-Rice singlet"~\cite{Zhang1988} band
of the original three-band Hubbard model.
However,
the approach that a simpler one-band Hubbard model or $ t $-$ J $ models
are capable of describing correctly
the doped CuO$_2 $ planes
has been challenged recently~\cite{Adolphs2016}
and
we agree with this particular point of view
that a strongly correlated single-band model cannot provide
the appropriate description for the cuprates.
Indeed, our starting point for the description of the CuO$_2 $ planes
is the spin-fermion model,
which has been also extensively used
to describe the high-$ T_c $ superconductors~\cite{Kotliar1990}

Henceforth,
consider a single CuO$_2$ plane
containing localized spins located at the sites of a square lattice,
which is the appropriate topology for the cuprates.
The spin degrees of freedom are modelled
by the spin 1/2 AF Heisenberg model,
$
H_{\rm H }
=
J \sum_{ <ij> } {\bf S}_i \cdot {\bf S}_j
$,
where $ {\bf S}_i $ is the localized spin operator.
As the system is doped, charge carriers are bumped into the planes
and the localized spins interact with the
spin degrees freedom of the itinerant fermionic charge carriers
via a Kondo coupling,
$
H_{ \rm K }
=
J_{K} \sum_{ i }
{\bf S}_i \cdot {\bf s}_i
$,
where
$ {\bf s}_i
=
\sum_{ \alpha, \beta }
c^{ \dagger }_{ i \alpha }\vec{\sigma}_{\alpha\beta} c_{ i \beta} $
denotes the spin operator of an itinerant charge carrier,
which is written in terms of the Pauli matrices
$ \vec{\sigma} = ( \sigma_x , \sigma_y , \sigma_z ) $
and $ c^{\dagger}_{i\alpha} $ denotes the creation operator for a charge carrier at site $ i $
with spin $ \alpha = \uparrow, \downarrow $.
Combining the Heisenberg model, the Kondo coupling
and the kinetic term associated to the itinerant charge carries,
one obtain the spin-fermion model.

We formulate this model in the continuum limit,
by employing the spin coherent states. This amounts to
replacing the localized spin operators by
$ S {\bf N }( {\bf x } ) $,
where $ S $ is the spin quantum number
and  $  {\bf N }( {\bf x } )  $ is a classical vector
such that $ | {\bf N }( {\bf x } ) |^{ 2 } = 1 $.
$ {\bf N }$ is then decomposed into two perpendicular components,
${ \bf L}$ and ${ \bf n}$ ($ {\bf L } \cdot { \bf n} = 0 $),
associated respectively with ferromagnetic and antiferromagnetic fluctuations~\cite{Sachdev,Tsvelik95}.
In the continuum limit, where the lattice spacing should be very small,
$ { \bf N } ( { \bf x }) $ is decomposed as
$
{\bf N }( {\bf x } ) = a^2 { \bf L}( {\bf x } ) + ( - 1 )^{ | {\bf x} | }  \, { \bf n}( {\bf x } ) + O(a^4)
$,
where  $ a $ denotes the lattice parameter
and we also have $ | {\bf n }( {\bf x } ) |^{ 2 } = 1 $.

The continuum limit of the AF Heisenberg model
in the square lattice is the well-known nonlinear sigma model
(NLSM)~\cite{Manousakis1991},
which is given by the following density Hamiltonian,
\begin{equation}
{\mathcal H }_{ \rm H }
=
\frac{1}{2}
\left( \,
\rho_{ s } | {\bf \nabla } {\bf n } |^{ 2 } + \chi_{ \perp } S^{2 } | {\bf L } |^{ 2 }
\, \right)
+
i S {\bf L } \cdot \left(  {\bf n } \times \partial_{ \tau } { \bf n}  \right)
\, ,
\label{EqHNLSM}
\end{equation}
where $ \rho_{ s } = J S^{ 2 } $ is the spin stiffness,
$ \chi_{ \perp } = 4 J a^2$ is the transverse susceptibility
and the last term in the rhs of the above expression describes the Berry phase.

On the same token,
the hamiltonian density of the Kondo interaction becomes
\begin{equation}
{\mathcal H }_{ \rm K }
=
J_{ \mbox{\scriptsize{K}} } S
\;
{\bf L }
\cdot
\sum_{ \alpha, \beta }
\psi^{ \dagger }_{ \alpha } \left(  \vec{ \sigma } \right)_{ \alpha \beta } \psi_{ \beta }
\, ,
\label{EqHK}
\end{equation}
where the continuum fermion field $ \psi_{ \alpha }( {\bf x } ) $
corresponds to $c_{i\alpha}$.
Also notice that the oscillating contribution
from the antiferromagnetic fluctuations
cancels out as we integrate it over space~\cite{Tsvelik95}.

For the cuprates, it is well known that Dirac points appear
in the intersection of the nodes of the $ d $-wave superconducting gap
and the two-dimensional (2D) Fermi surface.
In that case, the quasiparticles dispersion exhibit
a Dirac-like linear energy dispersion~\cite{Wehling2014}.
Presently, we assume that
the dispersion of the charge carriers can be linearized
close to the Fermi surface
and therefore the carrier kinematics
in the continuum limit
is described by the Dirac-Weyl density hamiltonian,
as previously seen in \cite{Dirac-like},
\begin{equation}
\mathcal{H}_0
=
\psi^{\dagger}
\left(
i \,  \hbar v_{F}  \vec{\sigma}\cdot \vec{\nabla} - \mu
\right)
\psi
\, ,
\label{EqH0}
\end{equation}
where $ \psi^{\dagger }_{\sigma }$ has spinorial components
$ \psi^{\dagger }_{\sigma }
=
( \psi^{\dagger }_{1 \sigma }, \psi^{\dagger }_{2 \sigma } ) $
and $ v_F $ is the Fermi velocity.
The indices $ 1 $ and $ 2 $ denote odd and even lattice sites respectively.
Notice that the chemical potential $ \mu $ controls the total number of charge carriers
that are added to the itinerant band as the system is doped.

Hence, in the continuum limit
we may express the partition function of the spin-fermion model
as the following functional integral in the complex time representation
\begin{eqnarray}
\mathcal{Z}
& = &
\int
\mathcal{D}\psi
\,
\mathcal{D}\psi^{ \dagger}
\,
\mathcal{D}{\bf L}
\,
\mathcal{D}{\bf n}
\,
\delta\left( | {\bf n }  |^{ 2 } - 1 \right)
 \nonumber \\
& &
\hspace{-0.4cm}
\times
\exp
\left[ - \int_0^\beta d\tau \int d^2 x \,
\left( \,
{\mathcal H }
-
\psi^{ \dagger } i\partial_{ \tau } \psi
\, \right)
\right]
\, ,
\label{EqZ}
\end{eqnarray}
where
$ \beta = 1 / k_B T $, with $ k_B $ denoting the Boltzmann's constant
and $ T $ the system temperature,
and
$
{\mathcal H }
=
{\mathcal H }_{ \rm H }
+
{\mathcal H }_{ \rm K }
+
{\mathcal H }_{ 0 }
$,
is given by (\ref{EqHNLSM}), (\ref{EqHK}) and (\ref{EqH0}) respectively.

\section{The effective nonlinear sigma model}
\label{EffNLSM}

We start by Fourier transforming our model Hamiltonian
assuming that the ferromagnetic component of the vector spin
$ {\bf L } $ is small and approximately constant in space,
since we investigate the system in the long range AF state ordering.
Therefore, the Kondo interaction from (\ref{EqHK}) is approximated by
\begin{eqnarray}
H_{ \rm K }
& = &
J_{ \rm K } S
\,
\int d^{2} k \int d^{2} k'
\\ \nonumber
& &
\hspace{-1.0cm}
\times
\sum_{ \alpha, \beta }
\psi^{ \dagger }_{ \alpha } ({\bf k })
\left[
\int d^2 x
\;
e^{ - i \left( {\bf k } - {\bf k }' \right) \cdot {\bf x} }
\,
{\bf L }
\right]
\cdot
\left(  \vec{ \sigma } \right)_{ \alpha \beta }
 \psi_{ \beta } ({\bf k }')
\nonumber \\
& \approx &
\hspace{-0.3cm}
J_{ \rm K } S
\,
{\bf L }
\cdot
\int d^{2} k
\left[
\sum_{ \alpha, \beta }
\psi^{ \dagger }_{ \alpha } ({\bf k })
\left(  \vec{ \sigma } \right)_{ \alpha \beta }
 \psi_{ \beta } ({\bf k })
 \right]
\, .
\label{EqHKFT}
\end{eqnarray}
In this approximation,
we may introduce the Nambu field
$ \Phi^{\dagger} =
\left( \,
\psi^{\dagger}_{1,\uparrow} \,
\psi^{\dagger}_{2,\uparrow} \,
\psi^{\dagger}_{1,\downarrow} \,
\psi^{\dagger}_{2,\downarrow} \,
\, \right) $
in order to express the fermionic part
of our model Hamiltonian
$
\mathcal{H}_{ \psi }
\equiv
\mathcal{H}_{ \rm K }
+
\mathcal{H}_0
-
\psi^{ \dagger } i\partial_{ \tau } \psi
$
in momentum space, $ {\bf p } = \hbar {\bf k } $,  as
$
\mathcal{H}_{ \psi }
=
\Phi^{\dagger}({\bf k }) \, \mathcal{A} \, \Phi({\bf k })
$,
where the matrix $ \mathcal{A} $ above is given by
\begin{equation}
\mathcal{A}=
\left(
 \begin{array}{cccc}
\tilde{ \mu }_{ + } & k_{-} &L_{ - } & 0\\
 k_{+} &  \tilde{ \mu }_{ + } & 0 & L_{ - } \\
 L_{ + }& 0 & \tilde{ \mu }_{ - } & k_{-} \\
 0 & L_{ + } & k_{+} & \tilde{ \mu }_{ - }
\end{array}
\right)
\, ,
\label{EqA}
\end{equation}
with the following definitions
in the above expression,
$ k_{ \pm } = - \hbar v_F ( k_x \pm i  k_y ) $,
$ \tilde{ \mu }_{ \pm } = - i \omega_{n} + \mu \pm J_K S \, L_z $
and
$ L_{ \pm } = J_K S \, \left( L_{ x } \pm i L_{ y }  \right) $.

We may now integrate exactly
the fermionic contribution of the partition function in (\ref{EqZ}),
which is a simple Gaussian path integral and, hence, proportional to
$ \det \mathcal{A} $.
Therefore,
\begin{eqnarray}
\ln \mathcal{Z}_{ \psi }
& = &
\ln\left( \prod_{ {\bf p }, n }\det \mathcal{A} \right)
\nonumber \\
& = &
\int d^2 k
\, \sum_{n}
\ln
\left[
(\hbar v_{F} k)^{2}-\left(
\mu_{-}-i \omega_{n}
\right)^2
\right]
\nonumber \\
& &
\hspace{-1.cm}
+
\int d^2 k
\, \sum_{n}
\ln
\left[
(\hbar v_{F} k)^{2}-\left(
\mu_{+}-i \omega_{n}
\right)^2
\right]
\, ,
\label{EqLnZPsi}
\end{eqnarray}
where
$ \mu_{\pm}=\mu \pm | J_{ K }\bf{ L } |$
and $ \omega_{ n } = ( 2 n + 1) \pi \beta^{-1} $
are the Matsubara frequencies for fermions.
Performing the sum over $ \omega_{ n } $
and after some algebra we get
\begin{eqnarray}
\hspace{-0.4cm}
\ln \mathcal{Z}_{ \psi }
& = &
\int d^2k
\,
\sum_{s = \pm 1}
\left\{
\beta \hbar v_{F} k
+
\ln
\left[
1 + e^{-\beta \left( \hbar v_{F} k + \mu_{s} \right) }
\right]
\right.
\nonumber \\
& &
\left.
+
\ln
\left[
1 + e^{-\beta \left( \hbar v_{F} k - \mu_{s} \right) }
\right]
\right\}
\, ,
\label{EqLnZPsi2}
\end{eqnarray}
which is the same result obtained
for a noninteracting relativistic system with a Zeeman term applied to it
(e.g.~\cite{Heron}),
but presently with $ | J_{ K } \bf{ L } | $
corresponding to an external ``magnetic field".
Furthermore, notice that (\ref{EqLnZPsi2})
yields to the partition function of a free fermion system
when $ { \bf L } \rightarrow  0 $,
as should be expected.

Now, we can integrate the above expression
over the first Brillouin zone in momentum space.
Also, notice that the Fermi surface of the YBCO system
has rotational symmetry around the point $(\pi,\pi)$.
Using translational invariance, we shift the momentum around this point,
thereby simplifying the integration over the first Brillouin zone.
Introducing the change of variable $ y = a_D k / \pi $,
where $ a_D $ is the lattice spacing between dopants,
we get
\begin{eqnarray}
\ln \mathcal{Z}_{ \psi }
& = &
\frac{ \pi }{ 3 } \gamma
+
\frac{ \pi }{ 2 }
\,
\sum_{ s = \pm 1 }
\int_{ 0 }^{ 1 } dy \, y
\,
\left\{
\ln
\left( 1 + e^{-\gamma  y } z_{ s } \right)
\right.
\nonumber \\
& &
\left.
+
\ln
\left( 1 + e^{-\gamma  y } z_{ s }^{ -1 } \right)
\right\}
\, ,
\label{EqLnZPsi3}
\end{eqnarray}
where we have introduced the dimensionless parameters
$ \gamma = \beta \, \hbar v_F \pi / a_D  $
and $ z_{ s } = e^{ - \beta \mu_{ s } } $,
with $ s = \pm 1 $.
Moreover, since we have assumed that $ | {\bf L } | $ is small,
we can expand $ \mathcal{Z}_{ \psi } $
as a Taylor series,
\begin{equation}
\mathcal{Z}_{ \psi }
=
\exp
\left[ \,
A( T ,\mu ) + B( T , \mu ) \, | {\bf L } |^2 + O\left(\, | { \bf L } |^4 \, \right)
\, \right]
\, ,
\label{EqZSeries}
\end{equation}
where
the first factor in the rhs of the above expression,
$ \exp\left[ A\left( T, \mu, \gamma \right) \right] $,
does not contribute to the effective NLSM
that will be obtained at the end of this section
or to the calculation of the N\'eel temperature in the next section
and hence shall be neglected.
On the other hand,
$ B( T , \mu ) $ seen in (\ref{EqZSeries})
is given by
\begin{eqnarray}
\hspace{-0.7cm}
B( T , \mu )
& = &
\frac{ \pi }{ 2 }
\left( \frac{ J_K S }{ \gamma } \right)^2
\left(
\beta \gamma
\left[
1 +
\frac{ \sinh\left(\beta \gamma \right) }
{ \cosh\left(\beta \gamma \right) +  \cosh\left(\beta \mu \right)}
\right]
\right.
\nonumber \\
& &
\hspace{-0.75cm}
\left.
+
\sum_{ s = \pm 1 }
\left\{
\ln\left( 1 + e^{ s \beta \mu } \right)
+
\ln\left[ 1 + e^{ \beta \left( \gamma + s \mu \right) } \right]
\right\}
\right)
\, .
\label{EqB}
\end{eqnarray}

Combining (\ref{EqB}) and (\ref{EqHNLSM}),
the NLSM becomes
\begin{eqnarray}
\tilde{ {\mathcal H } }_{ \rm H }
& = &
\frac{1}{2}
\left[ \,
\rho_{ s } | {\bf \nabla } {\bf n } |^{ 2 }
+
\tilde{ \chi }_{ \perp }\left( T, \mu \right) \, | {\bf L } |^{ 2 }
\, \right]
\nonumber \\
& &
+
i S {\bf L } \cdot \left(  {\bf n } \times \partial_{ \tau } { \bf n}  \right)
\, ,
\label{EqEffHNLSM}
\end{eqnarray}
where the above new transverse susceptibility
$
\tilde{ \chi }_{ \perp }\left( T, \mu \right)
=
\chi_{ \perp } S^2 - \frac{ 2 }{ \beta } B\left( \mu, T \right)
$
includes the effect of doping,
since it depends on the chemical potential.

Inserting the expression for
$ \tilde{ \chi }_{ \perp } $
in (\ref{EqZ})
and integrating over $ {\bf L } $
we finally get
the partition function, except for a multiplicative factor,
\begin{eqnarray}
\hspace{-0.75cm}
\mathcal{Z}
=
\int
\mathcal{D}{\bf n}
\,
\delta\left( | {\bf n }  |^{ 2 } - 1 \right)
\exp
\left(
- \int_0^{ \beta } d\tau
\int d^2 x
\,
\tilde{ \mathcal{ H } }_{ \rm eff }
\right)
\, ,
\label{EqZ2}
\end{eqnarray}
where the effective NLSM is
\begin{equation}
\tilde{ \mathcal{ H } }_{ \rm eff }
=
\frac{ \rho_{ s } }{2}
\left[ \,
| {\bf \nabla } {\bf n } |^{ 2 }
+
\frac{ 1 }{ c( T, \mu )^2 }
\,
| \partial_{ \tau }{\bf n } |^{ 2 }
\, \right]
\, ,
\label{EqEffHNLSM2}
\end{equation}
with the new spin wave velocity given by
\begin{equation}
c( T, \mu )
=
\sqrt{
\rho_s \, \chi_{ \perp } \,
\left[
1
-
8 \frac{  B\left( \mu, T \right) }{ \beta  \chi_{ \perp } }
\right]
}
\, .
\label{Eqc}
\end{equation}
Notice that $ c $ in the above expression
reduces to $ c = \sqrt{ \rho_s \, \chi_{ \perp } } $
in the absence of the Kondo coupling, as should be expected.

In the present approach
the spin wave velocity is finite for $ \mu = 0 $,
which is related to the parent compounds of the cuprates
and it is an even function with respect to the chemical potential.
Indeed, positive values of $ \mu $ are related to hole-doped cuprate superconductors,
while $ \mu < 0 $ corresponds to the electron-doped compounds.
Notice that doped electrons enter the Cu sites for the electron-doped compounds,
which is not the case for the hole-doped cuprates,
where doping introduces carriers at the oxygen sites.
However, in a continuum limit description
there is no difference between holes moving on an otherwise inert O-lattice,
in the presence of an AF background on the Cu sites
and electrons moving on the Cu sites
in the presence of the same AF background and an inert O-lattice.
The only difference perhaps would be on the value of the exchange coupling.

\section{N\'eel temperature calculation and comparison with experimental data}
\label{Experimental}

We start our analysis pointing out
that the Coleman-Mermin-Wagner-Hohenberg theorem
prevents the appearance of a long-range magnetic order at finite temperatures
for any 2D system~\cite{MermimWagner1966}.
Therefore, we add an small out-of-plane interlayer coupling
$ J_{ \perp } $,
so that the partition function for a stack of CuO$_2 $ planes
labeled by the subscript $ i $ becomes
\begin{eqnarray}
\mathcal{Z}
& = &
\prod_i
\int
\mathcal{D}{\bf n}_i
\,
\exp\left\{
- \int_0^{ \hbar \beta } d\tau
\int d^2 x
\right.
\nonumber \\
& &
\hspace{-0.6cm}
\left.
\times
\delta\left( | {\bf n }_i  |^{ 2 } - 1 \right)
\left[
\frac{ \tilde{ \mathcal{ H } }_{ \rm eff } }{ \hbar }
+
\frac{ \rho_s \alpha }{ \hbar }
\left( {\bf n }_{ i + 1} - {\bf n }_i \right)^2
\right]
\right\}
\, ,
\label{EqZeff2}
\end{eqnarray}
where we have introduced the parameter
$ \alpha = \left( 1 / a^2 \right) J_{ \perp } / J $.

Assuming that there is an external magnetic field
applied to the system,
one may calculate $ T_N $
with several approaches,
among them spin-wave theories (SWT)
and field-theoretical calculations,
which takes into account the contribution of the spin-fluctuation excitations
(neglected in the SWT).
Both the standard and self-consistent SWT
are shown to be insufficient to quantitatively describe the experimental data
for the parent compounds of the cuprates superconductors,
while the results calculated
for a large $ N $ expansion are in good agreement
with the experiments~\cite{Katanin2007}.
Hence,
we take the expression
for $ T_N $ obtained to order $ 1/N $ in a large $ N $ expansion
from the effective model in (\ref{EqZeff2}),
which is given by~\cite{Katanin1997}
\begin{equation}
T_N
=
4 \pi \rho_s
\left[
\ln
\left(
\frac{ 2  \, T_N^2  }{ \alpha \left( \hbar c' \right)^2 }
\right)
+
3 \ln
\left(
\frac{ 4 \pi \rho_s  }{ T_N }
\right)
-0.0660
\right]^{-1}
\, ,
\label{EqTN}
\end{equation}
where $ c' = \left( a / \hbar \right)  c $.
Moreover, we have set $ k_B = 1 $ for the sake of simplicity.
Notice that the above expression provides a self-consistent equation
for the calculation of the N\'eel temperature as a function of the chemical potential,
since the spin wave velocity $ c $ in (\ref{Eqc})
is expressed in terms of $ \mu $.

In the remaining of this section
we compare the results of the N\'eel temperature as a function of doping
with the available experimental data for YBCO.
Indeed, this compound has an almost circular shape for the Fermi surface
centered at $(\pi,\pi) $ in the reciprocal space~\cite{Hossain2008}
and the low energy dispersion can be approximated
to a linear relation
in the vicinity of the Fermi level,
which is consistent with the kinetic term given by (\ref{EqH0}).

In terms of the three-band Hubbard model parameters,
the exchange coupling between the Cu magnetic moments
is given by~\cite{Zannem1988,Kampf1994}
\begin{equation}
J
=
\frac{4 t_{pd}^{4}
}{
\left(\Delta E + U_{pd}\right)^{2}
}
\left(
\frac{1}{ U_{d} } + \frac{ 2 }{ 2 \Delta E + U_{p} }
\right)
\label{Jd}
\end{equation}
and the Kondo coupling of an itinerant oxygen hole spin
and the nearest local Cu spin is~\cite{Zannem1988,Kampf1994}
\begin{equation}
J_{k}
=
t_{pd}^{2}
\left(
\frac{1}{ \Delta E } + \frac{1}{ U_{d} - \Delta E }
\right)
\, .
\label{JK}
\end{equation}
Estimates for the microscopic parameters of the three-band model Hamiltonian
have been obtained from local-density functional techniques~\cite{McMahan1990}:
$ t_{pd}= 1.3 $, $ U_{d} = 7.3 $, $ U_p = 5 $, $ U_{pd} = 0.87 $
and $ \Delta E = 3.5 $,
all given in units of eV.
Inserting the above parameters values in (\ref{EqTN}) yields
$ T_N \approx 420 $ K for the undoped system ($ \mu = 0 $),
which is in excellent agreement with the experimental data for YBCO,
and therefore we employ those values in our numerical calculation from now own.

\begin{figure}
[ht]
\centerline
{
\includegraphics[
clip,
angle=90,
width=1.\columnwidth]
{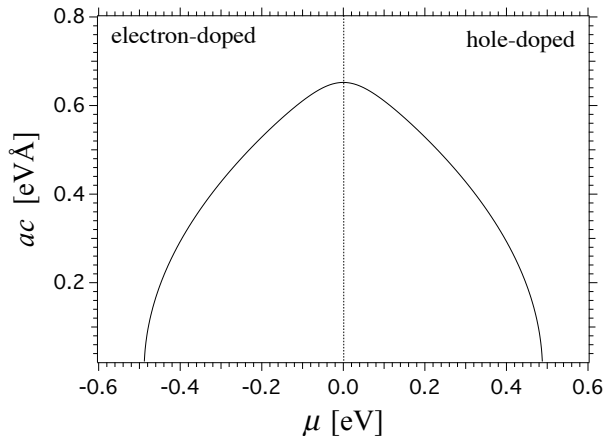}
}
\caption{Spin wave velocity multiplied by the lattice parameter $ a c $
as a function of the chemical potential $ \mu $ for $ T = 420 K $.}
\label{FigcXmu}
\end{figure}
Fom (\ref{Eqc}),
one can calculate  the spin wave velocity
as a function of the chemical potential numerically
for a particular temperature,
provided the values for
$ \rho_s $, $ \chi_{ \perp } $ and $ \gamma $ are given.
In our case, these are calculated from $ J_K $ and $ J $
in  (\ref{Jd}) and (\ref{JK}),
with the parameters values given above,
and also
$ {\hbar} v_{f}=1.15 $ eV \AA,
$ a_D  =  2.68 $ \AA \ \ and $ a = \sqrt{2} \,  a_D $~\cite{Marino2001}.
Our numerical results are shown in Fig.~\ref{FigcXmu}
for the particular value of $ T = 420 $ K.
Notice that for the undoped parent compound,
we assume that the Dirac point is at $ \mu = 0 $,
which is exactly the case for the cuprates.
As the chemical potential increases for positive values,
which means that holes are added to the CuO$_2 $ planes,
$ c $ vanishes indicating
the destruction of the long range AF order,
what is in agreement with the experimental results.
On the other hand,
for negative values of $ \mu $,
we also have that the spin wave velocity vanishes
as $ \mu $ increases in modulus,
which corresponds to the doping of electrons, instead of holes, to the system.
Therefore our approach provides a unified description for the phase diagrams
of the $ p $-type and $ n $-type cuprate superconductors,
where $ \mu > 0 $ are related to hole-doped systems,
while $ \mu < 0 $ are electron-doped ones.

\begin{figure}
[ht]
\centerline
{
\includegraphics[
clip,
angle=90,
width=1.\columnwidth]
{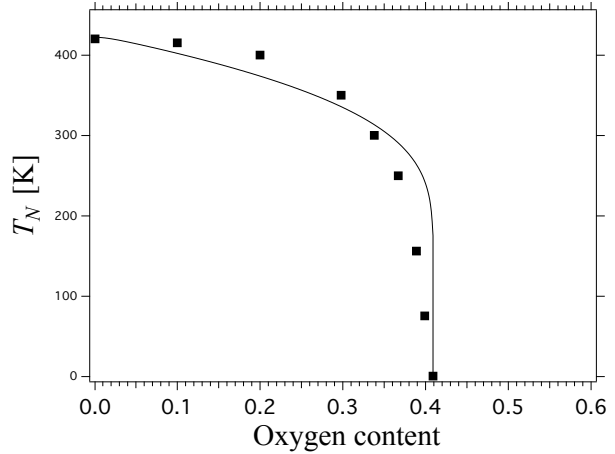}
}
\caption{N\'eel temperature $ T_N $ as a function of doping $ x $ for YBCO.
Squares indicate experimental data from Ref.~\cite{Rossat-Mignod1991}.
}
\label{FigTNXx}
\end{figure}
We can also calculate $ T_N $
as a function of the chemical potential from (\ref{EqTN}).
It is well known that the occupancy of charge carriers increases
as $ \mu > 0 $ also increases.
Therefore, in order to compare our numerical results
with the AF part of the phase diagram experimentally obtained for YBCO,
we follow a phenomenological approach
relating the chemical potential and the doping as
$ \mu - \mu_0 \propto \left( x - x_0 \right)^{ \beta } $,
where $ \mu_0 $ and $ x_0 $
are the chemical potential and the doping
for which $ T_N $ reaches zero respectively.
Since $ \mu_0 $ is numerically calculated
and $ x_0 $ is given by experiments,
the proportionality constant is uniquely defined and $ \beta $
is the single parameter which has been
adjusted to the available data .
The results of the AF phase diagram for the YBCO parameters
and $ \beta \approx 3 $
are shown in Fig.~\ref{FigTNXx}
and we see that our theory
is in good agreement with the experiments~\cite{Rybicki2016,Rossat-Mignod1991}.
Moreover, we also have that the magnetic order
is suppressed as charge carriers are added to the system,
as should be expected.

\section{Conclusions}
\label{conclusions}

Starting from the spin-fermion model
we have obtained an effective nonlinear sigma model
which takes into account the effects of doping in the system.
Taking the appropriate values of the parameters for YBCO,
we have calculated the AF phase diagram
as a function of the dopant concentration
and the results presented here
are in good agreement
with the experimental data~\cite{Rybicki2016,Rossat-Mignod1991}.
Notice that several studies indicate that there is a quantum phase transition
as the ratio $ J /J_{K } $ increases,
starting from a Fermi liquid state,
the system becomes a spin liquid~\cite{Bernard2011}.
Presently, on the other hand, the interaction couplings are not model parameters,
but provided by the experimental data available for the YBCO compound
and henceforth it is remarkable that the system presents an AF arrangement
in the absence of doping for the given values of $ J $ and $ J_{ K } $,
which is in agreement with the phenomenology of the cuprates.
Moreover, the calculated N\'eel temperature
is consistent with the experimental data for $ T_{N} $,
since the results were obtained without resorting to any kind of parameter adjustment.

Recently, also starting from the spin-fermion model,
we have obtained an effective interaction
among the charge carriers of the system,
which produces a dome-shaped SC high critical temperature
versus doping~\cite{AOP} plot
that qualitatively reproduces the SC phase diagram
experimentally observed.
Hence, combining our results,
the following picture emerges:
for the effective model of the localized spins presented here,
where the itinerant fermions have been integrated out,
we get the suppression of the magnetic order
as charge carriers are added to the system;
for the effective model of the itinerant fermionic fields,
where the localized magnetic moments
have been integrated out,
we have the appearance of a dome-shape SC critical temperature
with the addition of charge carriers~\cite{AOP}.
Therefore, we have a theory
where the AF order is suppressed
and the SC phase arises as charge carriers added to the system,
which is the phenomenology
observed for several strongly correlated electronic systems~\cite{Scalapino2012}.
However, the complete phenomenology of the cuprates,
including its strange metal behaviour
in the underdoped regime remain unexplained~\cite{Keimer2015}.

Furthermore,
our results for positive values of the chemical potential
are related to the addition of holes,
while negative values correspond
to the addition of electrons.
Therefore, our results provide a unified description
for the phase diagrams of the hole-doped and the electron doped compounds,
which is consistent with the data provided for the
$ p $-type and $ n $-type cuprate superconductors~\cite{Armitage2010}.
Notice that further studies are required
in order to address the quantitative differences
between the values for the ordering temperatures of the compounds,
since both $ T_c $ and $ T_N $ depend on the couplings
$ J $ and $ J_K $ of the spin-fermion model,
which are given in terms of the microscopic parameters
of the specific system under investigation.
Nevertheless, taking a single set of parameter values
and assuming that doping effects
produce only a weak dependence on $ J $ and $ J_K $,
the results presented
naturally lead to a qualitative symmetry
between hole and electron doped cuprates.

Also, it is worthy to mention
that the approach employed here
is not just restrained to the cuprates,
but might be applied to several other compounds,
since we have only assumed 2D spatial dimensionality
and the presence of a relativistic dispersion relation
for the itinerant fermionic fields.
Remarkably, the presence of Dirac electrons
have also been observed
for the cuprates
and other $ d $-wave superconductors~\cite{Wehling2014},
what might suggest that Dirac electrons
might play a relevant role
in some of those condensed matter systems.

\section{Acknowledgements}
\label{Acknowledgements}
E. C. Marino has been supported in part by CNPq and FAPERJ.

\bibliographystyle{elsarticle-num.bst}
\bibliography{natbib}

\end{document}